\documentclass[12pt]{article}
\usepackage{a4wide}
\usepackage{latexsym}
\usepackage{amsmath}
\usepackage{amsfonts}
\usepackage{cite}
\usepackage{float}

\usepackage{pslatex}
\usepackage[latin1]{inputenc}
\usepackage[OT2,T1]{fontenc}

\newcommand{\bq}{\begin{eqnarray}}
\newcommand{\eq}{\end{eqnarray}}
\newcommand{\eps}{\varepsilon}

\begin{document}

\thispagestyle{empty}

\begin{flushright}
  MITP/14-011
\end{flushright}

\vspace{1.5cm}

\begin{center}
  {\Large\bf Tutorial on loop integrals which need regularisation but yield finite results\\
  }
  \vspace{1cm}
  {\large Stefan Weinzierl\\
\vspace{2mm}
      {\small \em PRISMA Cluster of Excellence, Institut f{\"u}r Physik, }\\
      {\small \em Johannes Gutenberg-Universit{\"a}t Mainz,}\\
      {\small \em D - 55099 Mainz, Germany}\\
  } 
\end{center}

\vspace{2cm}

% abstract -------------------------------------------------------------------------
\begin{abstract}\noindent
  {
In this pedagogical note I 
will discuss one-loop integrals, where 
(i) different regions of the integration region lead to divergences
and
(ii) where these divergences cancel in the sum over all regions.
These integrals cannot be calculated without regularisation, 
in spite of the fact that they yield a finite result.
A typical example where such integrals occur is the decay $H \rightarrow \gamma \gamma$. 
   }
\end{abstract}

\vspace*{\fill}

% main text ------------------------------------------------------------------------
\newpage

% ----------------------------------------------------------------------------------
\section{Introduction}
\label{sect:intro}

The Higgs decay into two photons is an important decay channel due to its clean signature.
This decay is a loop-induced process. There are no tree contributions since the Higgs boson carries
no electrical charge.
The lowest order contribution is therefore given by one-loop diagrams, where the decay proceeds through a 
virtual $W$-boson or fermion loop.
These decays have been calculated long time ago by several groups \cite{Ellis:1975ap,Ioffe:1976sd,Shifman:1979eb,Rizzo:1979mf}.
The recent discovery of the Higgs boson brought a renewed interest in these calculations.
On the arxiv a few groups \cite{Gastmans:2011ks,Gastmans:2011wh,deGouvea:2013fca}
questioned the validity of the old one-loop calculations.
There have been excellent replies \cite{Shifman:2011ri,Marciano:2011gm,Jegerlehner:2011jm}, why the old results are correct.
The purpose of this note is not to add another argument in favour of the old results, but
to analyse in detail what goes wrong in the argumentation of ref.~\cite{Gastmans:2011ks,Gastmans:2011wh}
(Higgs decay through a $W$-boson loop) and in the argumentation of ref.~\cite{deGouvea:2013fca} (Higgs decay through a fermion loop).
It turns out that the flaw in the argument is related in both cases to the integral
\bq
\label{problematic_integral}
 I_{\mu\nu} 
 & = & 
 \int \frac{d^Dk}{i \pi^{\frac{D}{2}}}
 \; 
 \frac{4 k_\mu k_\nu - g_{\mu\nu}  k^2}{\left(k^2-m^2\right)^3}.
\eq
The correct treatment of this integral is certainly known to experts in the field of loop integrals. 
The aim of this note is to make the subtleties of this integral known to a wider audience.

The integral in eq.~(\ref{problematic_integral}) has been written down within dimensional regularisation.
Performing the integral one obtains a finite result.
The main point of this note is to explain that a finite result does not imply that the integral can be calculated without regularisation.
Eq.~(\ref{problematic_integral}) provides one of the simplest counter-examples.
The reason can be seen from the following ``Gedanken''-calculation:
Let us divide the integration region (i.e. the $D$-dimensional loop momentum space) into regions (for example a region can be defined by a region
of the $D$-dimensional solid angle with no constraints on the radial variable).
Then the integration over such a region would be (ultraviolet) divergent, if done in 4 space-time dimensions. We therefore need a regulator.
Dimensional regularisation is the standard choice for loop integrals, but other choice are also possible.
After integrating over the individual regions one sums up all partial results.
In this sum the divergent parts cancel and one obtains a finite result.

In other words, the integrand of eq.~(\ref{problematic_integral}) has singularities in different regions, which lead to divergences when integrated
in four space-time dimensions.
The integral in eq.~(\ref{problematic_integral}) has the additional property that the divergent parts cancel in the sum.
An even simpler example of this situation is the integral
\bq
 F & = & \int\limits_0^1 dx \; x^{\eps} \left(1-x\right)^{\eps} \left( x + \frac{1}{x} - \frac{1}{1-x} \right)
\eq
with $\eps>0$. The integrand has singularities at $x=0$ and $x=1$. The factor $x^\eps (1-x)^\eps$ acts as a regulator.
The integral is finite and yields 
\bq
 F & = &
 \frac{\Gamma\left(1+\eps\right)\Gamma\left(2+\eps\right)}{\Gamma\left(3+2\eps\right)}
 =
 \frac{1}{2} + {\cal O}\left(\eps\right).
\eq
However this does not imply that we can remove the regulator before integration and
\bq
\int\limits_0^1 dx \; \left( x + \frac{1}{x} - \frac{1}{1-x} \right)
\eq
is an ill-defined expression.

This note is organised as follows:
In section~\ref{standard_way} we present the standard calculation of the integral $I_{\mu\nu}$ within dimensional regularisation.
The possible pitfall within a four-dimensional calculation is described in section~\ref{pitfall}.
In section~\ref{integrability} we investigate the integrability of the integrand in $D$ dimensions and in four dimensions.
Finally, section~\ref{conclusions} contains the conclusions.

% ----------------------------------------------------------------------------------
\section{Dimensional regularisation}
\label{standard_way}

In this section we review the standard calculation of the integral
\bq
\label{integral_I_mu_nu}
 I_{\mu\nu} & = &
 \int \frac{d^Dk}{i \pi^{\frac{D}{2}}} 
 \;
 \frac{4 k_\mu k_\nu - g_{\mu\nu} k^2}{\left(k^2-m^2\right)^3}
\eq
within dimensional regularisation.
We set $D=4-2\eps$.
Let us analyse the individual parts corresponding to the two terms in the numerator.
The integral corresponding to the term $(-g_{\mu\nu}k^2)$ in the numerator is easily calculated and yields
\bq
\label{integral_g_mu_nu}
 -g_{\mu\nu} 
 \int \frac{d^Dk}{i \pi^{\frac{D}{2}}} 
 \frac{k^2}{\left(k^2-m^2\right)^3}
 & = &
 -
 g_{\mu\nu}
 \left( 1 - \frac{1}{2} \eps \right)
 \; \Gamma\left(\eps\right) \left(m^2\right)^{-\eps}.
\eq
Let us now analyse the integral corresponding to the term $4 k_\mu k_\nu$ in the numerator.
The integral
\bq
 \int \frac{d^Dk}{i \pi^{\frac{D}{2}}} 
 \frac{4 k_\mu k_\nu}{\left(k^2-m^2\right)^3}
\eq
is a tensor integral. It does not depend on any external momenta.
From $D$-dimensional Lorentz symmetry it follows that it has to be proportional
to the $D$-dimensional metric tensor $g_{\mu\nu}$ \cite{Passarino:1979jh}:
\bq
\label{integral_I1}
 \int \frac{d^Dk}{i \pi^{\frac{D}{2}}} 
 \frac{4 k_\mu k_\nu}{\left(k^2-m^2\right)^3}
 & = &
 g_{\mu\nu} I_1.
\eq
The integral $I_1$ is obtained by contracting eq.~(\ref{integral_I1}) 
with the $D$-dimensional inverse metric tensor $g^{\mu\nu}$.
Using $g_{\mu\nu}g^{\nu\mu}=D$ one finds
\bq
\label{integral_I1_2}
 I_1 & = & 
 \frac{4}{D} 
 \int \frac{d^Dk}{i \pi^{\frac{D}{2}}} 
 \frac{k^2}{\left(k^2-m^2\right)^3}.
\eq
The integral on the right-hand side of eq.~(\ref{integral_I1_2}) is now the same as in eq.~(\ref{integral_g_mu_nu})
and we obtain
\bq
\label{integral_k_mu_k_nu}
 \int \frac{d^Dk}{i \pi^{\frac{D}{2}}} 
 \frac{4 k_\mu k_\nu}{\left(k^2-m^2\right)^3}
 & = &
 g_{\mu\nu} \Gamma\left(\eps\right) \left(m^2\right)^{-\eps}.
\eq
Adding up both contributions we find that
\bq
\label{result_dim_reg}
 I_{\mu\nu} & = & 
 \frac{1}{2} g_{\mu\nu}
 \; \Gamma\left(1+\eps\right) \left(m^2\right)^{-\eps}
 =
 \frac{1}{2} g_{\mu\nu} + {\cal O}\left(\eps\right).
\eq
within conventional dimensional regularisation.
This is the standard result within conventional dimensional regularisation.
Note that this result is finite in the limit $\eps \rightarrow 0$. However we cannot set $\eps$ to zero before the integration.
In the calculation above we split the integral into two pieces.
The two pieces in eq.~(\ref{integral_g_mu_nu}) and eq.~(\ref{integral_k_mu_k_nu}) are both ultraviolet divergent and need regularisation.
In the sum the divergences cancel, resulting in a finite result.

In particular, the integral on the left-hand side of eq.~(\ref{integral_I1}) is ultraviolet divergent and requires
a regulator. Within conventional dimensional regularisation the numerator $4 k_\mu k_\nu$ is a tensor in $D$ dimensions,
and as a consequence, the metric tensor $g_{\mu\nu}$ on the right-hand side of eq.~(\ref{integral_I1})
is in $D$ dimensions as well. 
The trace of the $D$-dimensional metric tensor $g_\mu^{\;\;\nu}$ is $D$ and thus the pre-factor in the denominator
of eq.~(\ref{integral_I1_2}) is $D$ (and not $4$).

% ----------------------------------------------------------------------------------
\section{The pitfall}
\label{pitfall}

In this section we present the (wrong) argumentation of a four-dimensional calculation 
which leads to a contradiction with the result of dimensional regularisation given in eq.~(\ref{result_dim_reg}).
The wrong argumentation goes as follows:
\begin{enumerate}
\item The integral yields a finite result, therefore it can be calculated in four space-time dimensions.
\item The integral does not depend on any external momentum
and since we work in four space-time dimensions we can replace in the numerator
\bq
\label{wrong_substitution}
 k_\mu k_\nu & \rightarrow & \frac{1}{4} g_{\mu\nu}^{(4)} k^2,
\eq
where $g_{\mu\nu}^{(4)}$ is the metric tensor in four space-time dimensions.
The factor $4$ in the denominator follows from the trace of the metric tensor in four space-time dimensions:
$\mathrm{Tr} \; g_\mu^{(4)\;\nu}=4$.
\end{enumerate}
With the substitution in eq.~(\ref{wrong_substitution}) one concludes that the integral $I_{\mu\nu}$ yields zero, in contradiction
with the result of eq.~(\ref{result_dim_reg}).

What goes wrong in this argumentation? The argument $(2)$ is correct, if the integrand is an integrable function
in four space-time dimensions in the sense of measure theory.
However, argument $(1)$ is wrong: In four space-time dimensions the integrand is not an integrable function in the sense
of measure theory.
This will be shown in the next section.

% ----------------------------------------------------------------------------------
\section{Integrability}
\label{integrability}

In this section we investigate if the integrand is an integrable function.
We discuss the general $D$-dimensional case as well as the four-dimensional case.
We focus our attention on the $I_{00}$ component of $I_{\mu\nu}$.
The $I_{00}$ component has a high degree of symmetry and we can always perform $(D-2)$ integrations trivially,
leaving $2$ non-trivial integrations.
Note that we end up with $2$ non-trivial integrations, 
independently if we start from $D$ space-time dimensions or four space-time dimensions.
The integrand for this two-dimensional integration carries the dependence on the original dimension of space-time.
This integrand is integrable for $D<4$, however it is not integrable for $D=4$.

We look at the $\mu=0$ and $\nu=0$ component of $I_{\mu\nu}$:
\bq
 I_{00} & = &
 \int \frac{d^Dk}{i \pi^{\frac{D}{2}}}
 \; 
 \frac{4 k_0^2 - k^2}{\left(k^2-m^2\right)^3}.
\eq
We first perform the usual Wick rotation to Euclidean space (with the appropriate change of the integration contour and
the substitutions $k_0=iK_0$ for the time component and $k_i=K_i$ for the spatial components). Our integral becomes
\bq
 I_{00} & = &
 \int \frac{d^DK}{\pi^{\frac{D}{2}}} 
 \;
 \frac{4 K_0^2 -K^2}{\left(K^2+m^2\right)^3},
\eq
where $K^2 = K_0^2 + K_1^2 + ... + K_D^2$ denotes the squared Euclidean norm.
We then introduce generalised spherical coordinates ($K_0=K\cos\theta_1$, etc. ) and arrive at
\bq
 I_{00} & = &
 \pi^{-\frac{D}{2}}
 \int\limits_0^\infty dK K^{D-1}
 \int d\Omega_D
 \;
 \frac{K^2 \left( 4 \cos^2 \theta_1 -1 \right)}{\left(K^2+m^2\right)^3}.
\eq
$d\Omega_D$ is the measure for the solid angle in $D$ dimensions:
\bq
 \int d\Omega_{D} 
 & = & 
 \int\limits_{0}^{\pi} d\theta_{1} \sin^{D-2} \theta_{1}
 ... \int\limits_{0}^{\pi} d\theta_{D-2} \sin \theta_{D-2} 
 \int\limits_{0}^{2 \pi} d\theta_{D-1} 
 = 
 \frac{2 \pi^{D/2}}{\Gamma\left( \frac{D}{2} \right)}.
\eq
The integrations over $(D-2)$ of the $(D-1)$ angular variables are trivial.
Performing the integrations over $\theta_2$, $\theta_3$, ..., $\theta_{D-1}$
one obtains
\bq
\label{integral_2dim}
 I_{00} & = &
 \frac{2}{\sqrt{\pi} \Gamma\left( \frac{D-1}{2} \right)}
 \int\limits_0^\infty dK 
 \;
 \frac{K^{D+1}}{\left(K^2+m^2\right)^3}
 \;
 \int\limits_0^\pi d\theta_1 
 \;
 \sin^{D-2} \theta_{1}
 \left( 4 \cos^2 \theta_1 -1 \right).
\eq
Thus we managed to reduce the $D$ integrations to $2$ non-trivial integrations.
These two integrations clearly expose the problem.
The integrand of this two-dimensional integration carries the dependence on the original number of space-time dimensions $D$.
The integrand in eq.~(\ref{integral_2dim}) has the additional property that the integrand is factorised with respect to
the variables $K$ and $\theta_1$.
In $D<4$ space-time dimensions we can perform the integrations.
The radial integral yields
\bq
\label{radial_integration}
 \int\limits_0^\infty dK 
 \;
 \frac{K^{D+1}}{\left(K^2+m^2\right)^3}
 & = &
% \frac{1}{2} 
% \frac{D}{4-D}
% \left(m^2\right)^{\frac{D}{2}-2} \frac{\Gamma\left(\frac{D}{2}\right)\Gamma\left(3-\frac{D}{2}\right)}{\Gamma\left(3\right)}
% = 
 \frac{1}{2 \eps}
 \left(m^2\right)^{-\eps} \frac{\Gamma\left(3-\eps\right)\Gamma\left(1+\eps\right)}{\Gamma\left(3\right)}.
\eq
As before, we set $D=4-2\eps$.
This integral is convergent for $D<4$ (i.e. $\eps>0$), but divergent for $D=4$ (i.e. $\eps=0$).
Phrased differently, for $D=4$ the integrand is not an integrable function.
The divergence for $D=4$ is of ultraviolet origin.
The angular integration yields
\bq
\label{angular_integration}
 \int\limits_0^\pi d\theta_1 
 \;
 \sin^{D-2} \theta_{1}
 \left( 4 \cos^2 \theta_1 -1 \right)
 & = &
% \left(\frac{4}{D} -1 \right)
% \frac{\Gamma\left(\frac{1}{2}\right)\Gamma\left(\frac{D-1}{2}\right)}{\Gamma\left(\frac{D}{2}\right)}
% =
 \eps
 \frac{\Gamma\left(\frac{1}{2}\right)\Gamma\left(\frac{3}{2}-\eps\right)}{\Gamma\left(3-\eps\right)}.
\eq
This integration yields zero for $D=4$. However, for $D \neq 4$ we obtain a finite non-zero value.
The Taylor expansion of the $D$-dimensional result around $D=4$ starts at order $\eps^1$.
Combining the results in eq.~(\ref{radial_integration}) and in eq.~(\ref{angular_integration})
we obtain for $I_{00}$
\bq
 I_{00} & = &
 \frac{1}{2} \Gamma\left(1+\eps\right) \left( m^2 \right)^{-\eps}
 =
 \frac{1}{2} + {\cal O}\left(\eps\right),
\eq
in agreement with the previous result in eq.~(\ref{result_dim_reg}).
Let us emphasize that the finite non-zero value of $1/2$ for $I_{00}$ in the $\eps \rightarrow 0$ limit  
is obtained from the two integrations
as follows:
The radial integration is ultraviolet divergent and requires regularisation. In $D<4$ space-time
dimensions the Laurent series of the result starts at $1/\eps$. The pole $1/\eps$ reflects the ultraviolet
divergence.
The angular integration is always finite and has the additional property, that it vanishes in $D=4$
space-time dimensions.
The Taylor expansion of the result of the angular integration therefore starts with a term proportional
to $\eps$.
The product of these two results has therefore a Taylor expansion starting at $\eps^0$, resulting
in the non-vanishing value $1/2$ for $I_{00}$.

We can look at this result from a different perspective: 
Let us divide the integration interval $[0,\pi]$ for the variable $\theta_1$ into
$N$ segments $[\theta^{(i-1)},\theta^{(i)}]$ with $i \in \{1,2,...,N\}$ and
\bq
 0 = \theta^{(0)} < \theta^{(1)} < ... < \theta^{(N-1)} < \theta^{(N)} = \pi.
\eq
Then we can divide the $D$-dimensional loop momentum space in $N$ regions, where the $i$-th region
is defined by
\bq
 \theta_1 & \in & [\theta^{(i-1)},\theta^{(i)}],
\eq
and no additional constraints are imposed on the radial variable $K$ and on the remaining angular
variables $\theta_2$, ..., $\theta_{D-1}$.
In four space-time dimensions the integration over the $i$-th region is divergent 
and requires regularisation.
Using dimensional regularisation, the integration over the $i$-th region is convergent for $D<4$.
The result of the integration over each region is thus a Laurent series in $\eps$, starting at $1/\eps$.
Summing up the results from the $N$ regions, the $N$ coefficients of the $1/\eps$-terms add up to zero,
yielding a final result with a finite $\eps \rightarrow 0$ limit.
The integral in eq.(\ref{problematic_integral}) has therefore the property
that different regions of the integration region lead to divergences.
This implies that the integral needs a regulator.
The integral in eq.(\ref{problematic_integral}) has the additional property, that
the divergences cancel in the sum over all regions.
Therefore the integral yields a finite result, although it cannot be calculated without a regulator.

% ----------------------------------------------------------------------------------
\section{Conclusions}
\label{conclusions}

In this note we discussed in detail one particular one-loop integral.
This loop integral has in four space-time dimensions divergences in different corners of the integration region.
Once regulated, the sum over all divergent parts vanishes, leaving a finite result.
The main point of this note is to explain that this integral cannot be calculated without regularisation.
The discussed one-loop integral is relevant to the decay $H \rightarrow \gamma \gamma$ and the purpose of this note
is to make the subtleties of loop integration known to a wider audience.

% ----------------------------------------------------------------------------------
% references
\bibliography{/home/stefanw/notes/biblio}
\bibliographystyle{/home/stefanw/latex-style/h-physrev5}

\end{document}